\documentclass[preprint,12pt]{elsarticle}

\usepackage{amssymb}
\usepackage[cmex10]{amsmath}
\usepackage{epstopdf}
\usepackage{algorithm}
\usepackage{textcomp}
\usepackage{blkarray, bigstrut} %
\usepackage{blindtext}
\usepackage[center]{caption}
\usepackage[font=bf]{caption} 
\usepackage{subcaption}
\usepackage{multirow}
\usepackage{mathtools}

\usepackage[a4paper,bindingoffset=0.2in,%
            left=0.9in,right=0.9in,top=0.9in,bottom=0.9in,%
            footskip=.25in]{geometry}

\usepackage[T1]{fontenc}
\usepackage[utf8]{inputenc}
\usepackage{tabularx,ragged2e,booktabs,caption}
\newcolumntype{C}[1]{>{\Centering}m{#1}}

\newcommand\scalemath[2]{\scalebox{#1}{\mbox{\ensuremath{\displaystyle #2}}}}

\usepackage{algpseudocode}
\usepackage{array}

\journal{Physical Communication}

\begin{document}

\begin{frontmatter}

\title{Optimization Algorithms for Improving the Performance of Permutation Trellis Codes}

%
%
%

\author[1,2]{Oluwafemi Kolade \corref{cor1}}
\ead{femikolade@gmail.com}
\author[1]{Mulundumina Shimaponda-Nawa}
\author[3]{Daniel J.J. Versfeld} 
\author[1]{Ling Cheng}

\cortext[cor1]{Corresponding author}
\address[1]{Optical Communication Laboratory, School of Electrical and Information Engineering, University of the Witwatersrand, Johannesburg, South Africa.}
\address[2]{Sibanye-Stillwater Digital Mining Laboratory (DigiMine), Wits Mining Institute (WMI), University of the Witwatersrand, Johannesburg, South Africa.}
\address[3]{Department of Electrical and Electronic Engineering, Stellenbosch University, Stellensbosch, South Africa.}

\begin{abstract}

In this paper, soft-decision (SD) decoders of permutation trellis code (PTC) with $M$-ary frequency shift keying are designed using three optimization algorithms and presented in four decoding schemes. In a concatenated code such as PTC, the Viterbi decoder for the outer convolutional code provides maximum likelihood decoding. Hence, the error correction performance is dependent on the decoding scheme used for the inner code. Due to the structure of the encoder with the modulation scheme, the channel output can be interpreted as an assignment problem. SD decoding can then be designed accordingly, using the presented, low-complexity optimization-based schemes. The bit error rate (BER) performance of the schemes are simulated in an additive white Gaussian noise (AWGN) and powerline communication (PLC) channel. The complexities of the schemes are also presented. The performance of the SD schemes are compared with the existing SD threshold detector, with BER results showing significant coding gain for certain codebooks. From the results, a reasonable trade-off between the complexity and coding gain is observed for a noisy channel such as the PLC channel.

\end{abstract}


\end{frontmatter}

\section{Introduction}

The error correction codes used in improving data communication over noisy channels require low-complexity, near maximum likelihood (ML) decoding performance to ensure practicality of use in their target communication systems. Permutation trellis code (PTC), which incorporates the serial concatenation of a permutation code and convolutional code is one of such codes, which has been proposed to improve communication in the harsh powerline channel and cognitive radio networks. 
The permutation property resulting from the combination of PTC with $M$-ary frequency shift keying ($M$-FSK) provides frequency spreading, which mitigates the effect of narrowband interference (NBI) and impulse noise (IN) in a PLC channel \cite{ferreira-vinck, ferreira-vinck-swart-beer, vinck, lukusa-ouahada-ferreira}. In cognitive radio networks, PTC with $M$-FSK in \cite{bardan-masazade-ozdemir-varshney, bardan-masazade-ozdemir-han-varshney}, finds an appropriate mapping in order to increase the data rate of secondary users at low power in the presence of NBI by primary users. In addition, the ability of PTC to mitigate the effect of primary user interference is shown. In another study, the use of PTC in conjunction with quadrature amplitude modulation (QAM) and quadrature phase shift keying (QPSK) is reported \cite{gagnon-haccoun}. In these schemes, an appropriate mapping of the binary output of the convolutional encoder to a suitable permutation codebook increases the distance between the constellation points. At the output of the channel, a soft-decision (SD), threshold detector (TD) is used in \cite{ferreira-vinck-swart-beer} to improve the performance in the PLC channel, while \cite{ouahada} compares the performance of a range of thresholds for distance preserving PTC obtained from high order Galois fields.

In the construction of PTCs, a one-to-one mapping exists between the outputs of the outer convolutional code and the codewords of the inner permutation code. While several mappings are possible, the Hamming distance between any two binary sequences and their corresponding permutation codeword is preserved, increased or decreased \cite{ferreira-vinck, ferreira-vinck-swart-beer, chang-chen-klove-tsai}. The symbols in the non-binary codeword can then be modulated using $M$-FSK, as it enables non-coherent detection. The detection uses the square law or envelope detection (ED) to select the signal with the highest energy from $M$ correlation receivers, each having a pair of correlators for the in-phase and quadrature elements of the signal \cite{proakis}. The decoder of PTC's inner permutation code then finds the permutation codeword having the minimum Hamming distance with the received codeword. For large permutation codebooks, this operation becomes very complex.
In \cite{kolade-versfeld-van-wyk}, the error correction performance of the inner code of PTC is improved by exploiting the outputs of the channel in order to design an iterative permutation soft-decision decoder (PSDD). While the Viterbi algorithm \cite{viterbi} provides ML decoding of PTC, the overall error correction performance is limited by the inner code's error correction capability. In other words, the overall error correction performance is dependent on the decoding scheme used at the output of the channel. SD detection on the channel output improves the performance of coded schemes in general, but limited, low-complexity SD schemes exist for PTC. 

In this paper, the impact of optimization algorithms as SD decoders of PTC's inner permutation code with $M$-FSK is evaluated, in order to improve the overall error correction performance of PTC. In order to improve the overall performance at low decoding complexity, the error correction capability of the PSDD in \cite{kolade-versfeld-van-wyk} is used in decoding PTC. In addition, a novel SD decoder is designed using the branch and bound (BB) \cite{little-murty-sweeney-karel, ross-soland} algorithm. The presented SD decoders are then combined with the Viterbi decoder in four different schemes in order to improve the overall bit error rate (BER) performance. Results show that the additional step of converting the SD output to its binary equivalent before decoding with the Viterbi algorithm produces better BER performance for both decoders. The complexities of the decoders are also presented. In some cases, the SD decoder reduces the decoding complexity for large codebooks, while simultaneously improving the BER performance.


\section{System Model}
\subsection{Permutation Trellis Codes}

\begin{figure}
\centering\includegraphics[width=30pc]{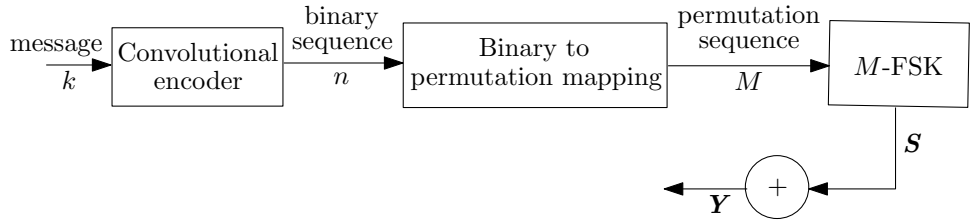}
\caption{System model for permutation trellis encoder.}
\label{fig:permutation-trellis-system-model}
\end{figure}

Permutation trellis codes combine convolutional codes and permutation codes at the encoder and are decoded using a modified Viterbi decoder \cite{ferreira-vinck-swart-beer}. A convolutional code is characterized by its rate $R_{\text{C}} = k/n$, the constraint length $K$ and the generator polynomial in an octal notation. The encoder uses $K-k$ shift registers and can be interpreted using a state transition diagram example shown in Fig. \ref{fig:permutation-trellis-system-model} having $2^{(K-k)}$ states. The octal generator notation is used to represent the XOR adder (output) connections of the inputs to the shift registers. As an example, an octal generator vector (7 5), having two elements represents a $k = 1$, $n = 2$ encoder which uses two ($n = 2$) XOR adders. There are one current input and two delayed inputs in the shift register. For the first adder (output), the binary vector of octal 7 is [1 1 1], while the binary vector of 5 is [1 0 1] for the second adder. A `1' indicates a connection between an input or a delayed input in the shift register and the XOR adder of the encoder, while a `0' indicates no connection to the adder. 

\begin{figure}[!ht] 
  \begin{subfigure}[b]{0.3\linewidth}
    \centering
    \includegraphics[width=10pc]{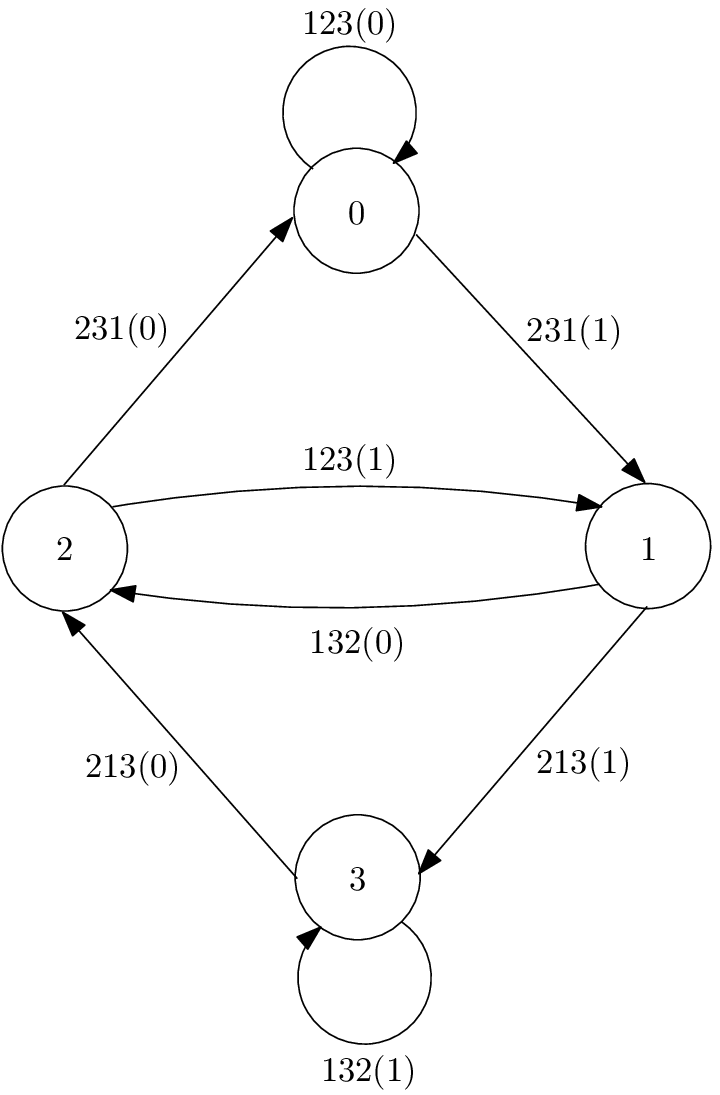}
    \caption{State transition diagram.} 
    \label{fig:state-transition-ptc}
    \vspace{2ex}
  \end{subfigure}
  \begin{subfigure}[b]{0.7\linewidth}
    \centering
    \includegraphics[width=24pc]{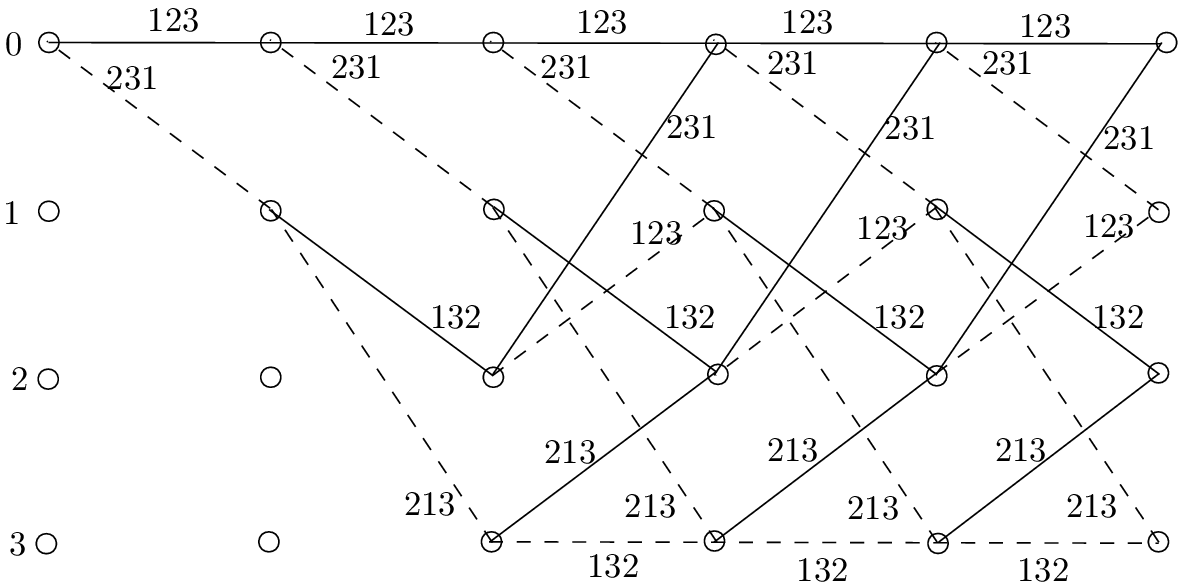}
    \caption{Trellis representation of the PTC.} 
    \label{fig:ptc-trellis-rate-1-2}
    \vspace{2ex}
  \end{subfigure}
  \caption{State transition diagram and equivalent trellis for $R_{\text{C}} = 1/2$ and $R_{\text{P}} = 1/3$.}
  \label{fig:state-transition-and-trellis-ptc} 
\end{figure}

The $n$-tuple output from the convolutional encoder is uniquely mapped onto a codeword belonging to a permutation codebook $\boldsymbol{C}$. The codebook $\boldsymbol{C}$ is a subset of all possible permutation of integers $c_1, c_2, \dots c_M$. All the codewords in $\boldsymbol{C}$ can be arranged in a matrix which consists of $2^n$ rows and each row contains a codeword $\boldsymbol{c}_q$ at row $q$ $(1 \leq q \leq 2^n)$. Any two codewords $\boldsymbol{c}_1, \boldsymbol{c}_2$ have a Hamming distance $d_{\text{H}}(\boldsymbol{c}_1, \boldsymbol{c}_2)$ between them, which is the number of positions where the two codewords differ. The permutation codebook has a minimum Hamming distance $d_{\text{min}}$, which is the lowest Hamming distance between all possible, distinct codeword pairs.
In mapping the convolutional codeword onto the permutation codeword, the combined code rate of PTC becomes $R_{\text{P}} = k/M$.
The permutation codebook is chosen such that $n \leq M$ and the mapping is done such that the Hamming distance between each binary sequence and its respective permutation sequence is preserved, reduced or increased. Two examples of $n = 2$ mapped onto an $M = 3$ codebook and $n = 3$ mapped onto an $M = 4$ codebook are shown in Table \ref{table:binary-to-permutation-mapping}. In Fig. \ref{fig:state-transition-ptc}, the state transition diagram of the $n = 2, M = 3$ mapping is shown, where the non-binary sequence replaces the binary output of the convolutional encoder as an input bit (in brackets) produces an output and transition between states. The trellis representation of the code is shown in Fig. \ref{fig:ptc-trellis-rate-1-2}, where the dotted lines represent an input bit `1', while the solid lines in the trellis represent an input bit `0'. 
\begin{table*}[t]
\caption{Two sample mappings of binary outputs of convolutional code to non-binary permutation codes}
{\begin{tabular}{cc|cc}\toprule
$n$ = 2 & $M$=3 & $n$ = 3 & $M$ = 4\\
\midrule
00 & 123 & 000 & 1234\\
\hline
01 & 132 & 001 & 1342\\
\hline
10 & 213 & 010 & 1423\\
\hline
11 & 231 & 011 & 2143\\ 
\hline
 & & 100 & 2314\\ 
\hline
 & & 101 & 2413\\ 
\hline
 & & 110 & 3241\\ 
\hline
 & & 111 & 3412\\ 
\hline
\end{tabular}}{}
\label{table:binary-to-permutation-mapping}\centering
\end{table*}
We refer the reader to other examples of distance increasing, reducing and preserving mappings in \cite{ferreira-vinck, ferreira-vinck-swart-beer, chang-chen-klove-tsai}.

\subsection{$M$-ary Frequency Shift Keying with PTC}

Consider a set of $M$ orthogonal signals $\boldsymbol{s}_{1}, \boldsymbol{s}_{2}, \dots, \boldsymbol{s}_{M}$, 
each having one of the $M$ different frequency components and a low pass vector representation \cite{proakis} 
\begin{equation}
    \begin{aligned}
        \boldsymbol{s}_{m} &= (\underbrace{0,\dots,0}_{m-1},\sqrt{E_s}, \underbrace{0, \dots, 0}_{M-m})^T.
    \end{aligned}
    \label{eq:mfsk-vector-representation}
\end{equation}
The notation $(\cdot)^T$ denotes the transpose operation and the symbol energy $E_s$ is in the $m$-th ($m = 1, 2, \dots, M$) position denoting the $m$-th frequency. The vector $\boldsymbol{c}_q$ is modulated by permuting the signal $\boldsymbol{s}_{m}$, $M$ times or over $M$ time slots. Each time slot corresponds to each integer in $\boldsymbol{c}_q$ and the value of each integer indicates the position of the symbol energy. Hence, an $M$-FSK modulated codeword is equivalent to a permutation matrix $\boldsymbol{S} = [s_{ij}] \in \{0, 1\}^{M \times M}$ of the form
\begin{equation}
  \begin{aligned} 
    \boldsymbol{S} = \left[ 
    \begin{array}{cccc}
    s_{11} & s_{12} & \dots & s_{1M} \\
    s_{21} & s_{22} & \dots & s_{2M} \\
    \vdots & \vdots & \ddots & \vdots \\
    s_{M1} & s_{M2} & \dots & s_{MM}
    \end{array} \right].
  \end{aligned} 
\end{equation}
The rows represent the $M$ available frequencies while the columns represent the time slots. At each time slot, the position $(i, j)$ is set to `1' if the frequency is conveying information, while other positions at the same time slot are set to `0'.

Assuming perfect synchronization between the transmitter and receiver, for a transmitted coded matrix $\boldsymbol{S}$, non-coherent detection over an AWGN and PLC channel are considered. In the PLC channel, NBI and IN are common noise sources \cite{middleton}. While IN may affect some or all frequencies in a time slot, NBI may affect one of the $M$ frequencies over a period of time as illustrated in \cite{ferreira-vinck-swart-beer}.
In the time domain, IN in an indoor environment occurs in short bursts and lasts for a short period of time. As a result, the occurrence may affect successive information-carrying symbols. In the case of NBI, the noise takes a frequency selective nature in which a certain frequency is affected over a period of time. Due to the random occurrence and short burst nature of IN, its arrival rate $\gamma$ in units per second is used to model the Poisson process. If each IN occurs at an average duration $T_{\text{noise}}$, then an average of $\gamma T_{\text{noise}}$ samples are affected by IN in 1 second (s) and $1 - \gamma T_{\text{noise}}$ samples are not affected by IN. At the output of the channel, each received sample can be modelled as
\begin{equation}
    y_{ij} = 
\begin{cases}
  
    e^{j\phi} s_{ij} + v_{\text{G}}, & \text{AWGN},\\
    e^{j\phi} s_{ij} + v_{\text{G}} + v_{\text{I}} \cdot p,              & \text{PLC}.
  \label{eq:awgn-equation-plc}
\end{cases}
\end{equation}
The phase $\phi$ is a random variable, distributed uniformly between 0 and $2\pi$, the noise sample $v_{\text{G}}$ has the Gaussian distribution $\mathcal{C}\mathcal{N} (0,\frac{N_0}{2})$ and $v_{\text{I}}$ models the IN as a complex variable with distribution $\mathcal{C}\mathcal{N} (0,N_i)$. The Poisson-distributed variable $p$ models the occurrence of the IN with an impulsive index $A = N_0/N_i$, describing the power of the impulse noise $N_i$ in terms of the Gaussian noise power $N_0$. 

The received code matrix $\boldsymbol{Y} = [y_{ij}] \in \mathcal{C}^{M \times M}$ is demodulated by comparing each column vector in $\boldsymbol{Y}$ received from the output of the channel. Hence, for every transmitted vector $\boldsymbol{s}_m$, the received vector $\boldsymbol{y}$ (which forms a column in $\boldsymbol{Y}$) is demodulated using ED by choosing the $\boldsymbol{s}_{m}$ with the highest envelope
\begin{equation} 
  \begin{aligned}
    r = \displaystyle\arg \max_{m \in M} |\boldsymbol{y}^H \cdot \boldsymbol{s}_{m}|.
  \end{aligned}
\label{eq:mfsk-envelope-detector}
\end{equation}
This produces an integer $r$ which is equivalent to the likely transmitted $m$-th frequency, 
hence a permutation of integers $\boldsymbol{r} = r_1, r_{2}, \dots, r_{M}$ per transmitted codeword. A more adequate threshold detector proposed in \cite{ferreira-vinck-swart-beer} sets a threshold of $\tau = 0.6\sqrt{E_s}$ on the received samples, such that 
\begin{equation}
    r_{ij} = 
\begin{cases}
    1, & \text{if } \bar{y}_{ij} \geq \tau,\\
    0,              & \text{otherwise},
    \label{eq:threshold-detector}
\end{cases}
\end{equation}
where $\bar{y}_{ij} = |y_{ij}|$. 
This results in an $M \times M$ matrix $\boldsymbol{R} = [r_{ij}] \in \{0,1\}^{M \times M}$, equivalent to a transmitted signal $\boldsymbol{S}$.

\subsection{Probability of Error of PTC}
Consider a set of transmitted code matrices $\boldsymbol{S}^1, \boldsymbol{S}^2, \dots, \boldsymbol{S}^V$ which are received over an AWGN channel and demodulated as $\boldsymbol{R}^1, \boldsymbol{R}^2, \dots, \boldsymbol{R}^V$. For each transmitted code matrix $\boldsymbol{S}^v$ ($v = 1,2,\dots,V$), received and demodulated as $\boldsymbol{R}^v$, the probability of erroneously receiving the $n$-tuple symbols encoded as $\boldsymbol{S}^v$ is given as \cite{bardan-masazade-ozdemir-varshney}

\begin{equation}
  \label{eq:pe-ptc-varshney}
    P_e = \frac{1}{M} \sum^{2^n}_{i=1} \sum^{2^{M^2}}_{v=1} P(\boldsymbol{D} = \boldsymbol{S^v} | \boldsymbol{R}^v) \cdot P(\boldsymbol{R}^v|\boldsymbol{S} = \boldsymbol{S^v}),
\end{equation}
where $P(\boldsymbol{R}^v|\boldsymbol{S} = \boldsymbol{S}^v)$ is the likelihood of receiving $\boldsymbol{R}^v$ if $\boldsymbol{S}^v$ is transmitted and is given as
\begin{equation}
  \label{eq:pe-ptc-varshney-probability-sent-received}
    P(\boldsymbol{R}^v|\boldsymbol{S} = \boldsymbol{S^v}) = \prod_{i=1}^{M}\prod_{j=1}^{M} \left( \prod_{w=1}^{W} P(\bar{y}_{ij} \geq \tau | s_{ij}) \right) \left( \prod_{u=1}^{U} P(\bar{y}_{ij} < \tau | s_{ij}) \right),
\end{equation}
where $U$ and $W$ are the number of zeros (`0's) and ones (`1's) respectively in $\boldsymbol{R}^v$.
$P(\boldsymbol{D} = \boldsymbol{S^v} | \boldsymbol{R}^v)$ is the probability of a correct decision because code matrix $\boldsymbol{D}$, which has the minimum Hamming distance with the received code matrix $\boldsymbol{R}^v$ is transmitted. The probabilities in (\ref{eq:pe-ptc-varshney}) and (\ref{eq:pe-ptc-varshney-probability-sent-received}), derived in \cite{bardan-masazade-ozdemir-varshney, bardan-masazade-ozdemir-han-varshney} are given as
\begin{equation}
  \label{eq:pe-ptc-varshney-probability-explained-11}
    P(\bar{y}_{ij} \geq \tau |s_{ij} = 1) = Q_1\left(\sqrt{2\frac{E_s}{N_0}}, 0.6\sqrt{2\frac{E_s}{N_0}} \right),
\end{equation}

\begin{equation}
  \label{eq:pe-ptc-varshney-probability-explained-01}
    P(\bar{y}_{ij} < \tau |s_{ij} = 1) = 1 -  P(\bar{y}_{ij} \geq \tau |s_{ij} = 1),
\end{equation}

\begin{equation}
  \label{eq:pe-ptc-varshney-probability-explained-00}
    P(\bar{y}_{ij} \geq \tau |s_{ij} = 0) = \text{exp} \left( -0.36\frac{E_s}{N_0} \right),
\end{equation}
and
\begin{equation}
  \label{eq:pe-ptc-varshney-probability-explained-10}
    P(\bar{y}_{ij} < \tau |s_{ij} = 0) = 1 - P(\bar{y}_{ij} \geq \tau |s_{ij} = 0).
\end{equation}
Note that the SNR per symbol $E_s/N_0$ is the ratio of the received signal energy $E_s$ to the noise power spectral density $N_0$, $Q_1(\alpha, \beta) = \int_{\beta}^{\infty} x \; \text{exp} \{ - \frac{x^2 + \alpha^2}{2} \} I_0(\alpha x) dx$ is the Marcum $Q$-function \cite{marcum} and $I_0(\alpha x)$ is the zeroth order of the modified Bessel function. The $P_e$ in (\ref{eq:pe-ptc-varshney}), henceforth referred to as the analytical hard decision (HD) decoder gives the probability of error for a transmitted symbol $n$. 
The convolutional code is an encoder with memory. Therefore, the decoder decodes the output of the demodulator as a chain of events whose length may not be fixed. In order to simplify the probability of error of the Viterbi decoder, the all-zero path is assumed to be transmitted. Therefore, the code's free distance $d_{\text{free}}$, which is the path which departs the all-zero path and first merges with the all-zero path is assumed to contain errors when compared with the transmitted all-zero path. As a result, the probability of error is bounded as \cite{proakis, lin-costello}

\begin{equation}
  \label{eq:pe-convolutional-code}
    P_{e, \text{CC}} \leq \sum_{d=d_{\text{free}}}^{\infty} a_dP_2(d),
\end{equation}
where $P_2(d)$ describes the probability of the decoded path having a Hamming distance of $d$ bits with the transmitted codeword. 
When the binary convolutional code is mapped onto the non-binary permutation codeword, the Viterbi HD decoder's $P_{e, CC}$ is modified as
\begin{equation}
  \label{eq:pe-ptc}
    P_{e, \text{CC}}^{\boldsymbol{c}} \leq \sum_{d=d_{\text{free}}^{'}}^{\infty} a_{d}^{\boldsymbol{c}}P_{2}^{\boldsymbol{c}}(d),
\end{equation}
where $a_{d}^{\boldsymbol{c}}$ and $P_{2}^{\boldsymbol{c}}(d)$ are similar to $a_d$ and $P_2(d)$ in (\ref{eq:pe-convolutional-code}) but correspond to the permutation codeword $\boldsymbol{c}_q$ mapped to a convolutional codeword and $d_{\text{free}}^{'}$ is the free distance of the non-binary code. Using the example mappings in Table \ref{table:binary-to-permutation-mapping}, the all-zero sequence is equivalent to $123$ for $M = 3$ and $1234$ for $M = 4$. Hence, $a_{d}^{\boldsymbol{c}}$ describes the paths with a Hamming distance of $d$ when compared with the transmitted permutation codeword equivalent to the corresponding all-zero codeword. If the path in error consists of $V$ stages, then a concatenation of matrices $\boldsymbol{R}^1, \boldsymbol{R}^2, \dots, \boldsymbol{R}^V$ are received for transmitted code matrices $\boldsymbol{S}^1, \boldsymbol{S}^2, \dots, \boldsymbol{S}^V$. Therefore, the probability $P_{2}^{\boldsymbol{c}}(d)$ is defined as
\begin{equation}
    P_{2}^{\boldsymbol{c}}(d) = 
    \begin{dcases}
        \sum_{d_{\text{H}}(\boldsymbol{S}^v, \boldsymbol{R}^v)={\frac{d}{2} + 1}}^{d} \prod_{v=1}^{V} P(\boldsymbol{R}^v | \boldsymbol{S}^v), \;\;\;\;\;\;\;\;\;\;\;\;\;\;\;\;\;\;\;\;\;\;\;\;\;\;\;\;\;\;\;\; \text{ odd } d,\\
        \sum_{d_{\text{H}}(\boldsymbol{S}^v, \boldsymbol{R}^v)={\frac{d}{2} + 1}}^{d} \prod_{v=1}^{V} P(\boldsymbol{R}^v | \boldsymbol{S}^v) + \frac{1}{2} \prod_{v=1}^{V} P(\boldsymbol{R}^v | \boldsymbol{S}^v), \;\;\;\;\;\;\;\;\; \text{even } d, \\
    \end{dcases}
\label{eq:definition-of-P2d}
\end{equation}
where $\boldsymbol{S}^v$ and $\boldsymbol{R}^v$ are the transmitted and received matrices respectively at each stage $v$ in the trellis. At each stage of the trellis, the decoder computes the metric
\begin{equation} 
  \begin{aligned}
    \label{eq:ptc-decoder-branch-metric}
    d_{\text{H}}(\hat{\boldsymbol{S}^v}, \boldsymbol{R}^v) = M - (\sum_{1 \leq i, j\leq M} (s_{ij} \land r_{ij})),
  \end{aligned}
\end{equation}
between the received matrix $\boldsymbol{R}^v$ and each branch $\hat{\boldsymbol{S}^v}$. The decoder chooses the branch with the smallest metric, where the notation $\land$ is the binary AND operation. As a result, at each stage $v$ of the trellis, the probability of error $P(\boldsymbol{R}^v | \boldsymbol{S}^v)$ can be evaluated as pairwise, hence a function of the difference between $\boldsymbol{R}^v$ and $ \boldsymbol{S}^v$ and is defined as
\begin{equation}
    P(\boldsymbol{R}^v | \boldsymbol{S}^v) = \frac{1}{2} \text{erfc}\left(\sqrt{\frac{R_{\text{P}}}{M}\frac{E_s}{N_0} d_{\text{H}}(\boldsymbol{S}^v, \boldsymbol{R}^v) }\right),
\label{eq:definition-of-P-r-t}
\end{equation}
where erfc($x$) is the complementary error function of $x$.
\subsection{The Permutation Soft-Decision Decoder}
As mentioned earlier, a permutation codeword of length $M$ permutes the $M$FSK signal in (\ref{eq:mfsk-vector-representation}), $M$ times. In each transmitted codeword matrix $\boldsymbol{S}$, all $M$ frequencies carry information such that each frequency is activated once in each row and column. When the received signal is affected by AWGN or PLC channel noise, the PSDD solves the general assignment problem by minimizing
\begin{equation} 
    Z_g = \sum_{i=1}^{M}\sum_{j=1}^{M} \hat{y}_{ij} s_{ij},
    \label{eq:optimization-problem}
\end{equation}
where $\hat{y}_{ij} = -|y_{ij}|$, $Z_g$ is the cost at iteration $g$ and (\ref{eq:optimization-problem}) is subject to
\begin{equation} 
    \label{eq:assignment-problem-explained}
    \begin{aligned} 
      \sum_{i=1}^{M} s_{ij} = \sum_{j=1}^{M} s_{ij} = \sqrt{E_s},
    \end{aligned}
\end{equation}
for $i = 1,2, \dots, M$, $j = 1,2, \dots, M$. The Hungarian algorithm (HA) \cite{kuhn} can find the matrix $\boldsymbol{S} = [s_{ij}]$ that produces the minimum cost $Z_1$ with a corresponding row-column representation of the permutation sequence 
\begin{equation} {\{(1, j_1), (2, j_2), \dots, (M, j_{M})}\}, \end{equation}
where $j_1, j_2, \dots, j_M$ is a permutation of integers $1, 2, \dots, M$. If the codeword corresponding to cost $Z_1$ produces a codeword $\boldsymbol{c} \notin \boldsymbol{C}$, then the PSDD proceeds to the next iteration $g = 2$. Using Murty's algorithm \cite{murty}, the PSDD ranks $Z_2, Z_3, \dots, Z_{2^n}$ in order to find a codeword $\boldsymbol{c} \in \boldsymbol{C}$. Using the solution matrix from $Z_1$, $n-1$ non-empty subsets of $\boldsymbol{Y}$ form nodes $N_1, N_2, \dots, N_{n-1}$, by partitioning $a_1$. Nodes in this case are defined as 
\begin{equation*} 
  \begin{aligned}
  \label{eq:murty-problem-1}
    N_1 = {\{\overline{(1, j_1)}}\}, \\ N_2 = {\{(1, j_1), (\overline{2,j_2)}}\}, 
  \end{aligned}
\end{equation*}

\begin{equation} 
  \begin{aligned}
  \label{eq:murty-problem-explained}
    \vdots
  \end{aligned}
\end{equation}

\begin{equation*} 
  \begin{aligned}
  \label{eq:murty-problem-2}
      N_{n-1} = {\{(1, j_1), \dots, \overline{(n-1, j_{n-1})}}\}.
  \end{aligned}
\end{equation*}

        
    
Row-column pairs with the bar ($\overline{\cdot,\cdot}$) indicate row-column elements to be replaced with $\infty$ while row-column elements without the bar are removed from $\boldsymbol{Y}$. The minimum cost is then solved for each node and the node with the least cost forms the next assignment $a_2$.

The PSDD makes an optimal decision (OD) by using a brute force method to find the codeword $\boldsymbol{c}$ having the highest cost among all possible permutations
\begin{equation} 
    Z_{\text{OD}} = \arg \min_{\boldsymbol{c}_{q} \in C} \sum_{i=1}^{M}\sum_{j=1}^{M} \hat{y}_{ij} s^{(\boldsymbol{c}_{q})}_{ij}, \quad \mbox{for} \ 1 \leq q \leq 2^n, 
   \label{eq:assignment-problem-solution-permutation-ml}
\end{equation}
where $s^{(\boldsymbol{c}_{q})}_{ij}$ is each element in the permutation matrix produced by a codeword $\boldsymbol{c}$.

\section{Proposed Soft-Decision Decoder and Schemes}

\subsection{Branch and Bound} \label{sec:branch-and-bound}
\begin{figure}
    \centering\includegraphics[width=20pc]{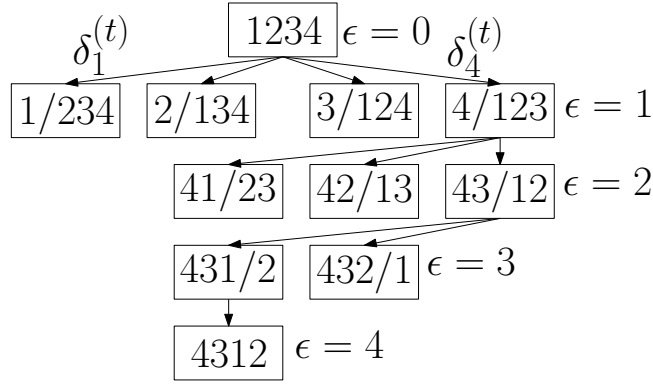}
    \caption{Tree-based method to decode permutation codes using branch and bound with $M = 4$.}
    \label{fig:branch-and-bound-tree-diagram}
\end{figure}

In order to solve $\boldsymbol{Y}$ using BB, a set of $M+1$ levels $\epsilon = 0, 1, 2, \dots, M$ are created and at each level $\epsilon > 0$, there exists nodes $t = 1, 2, \dots, M-\epsilon+1$ each with cost $\delta_{\epsilon}^{(t)} = \delta_{\epsilon}^{(1)}, \delta_{\epsilon}^{(2)}, \dots, \delta_{\epsilon}^{(M-\epsilon+1)}$ respectively. At $\epsilon = 0$, there exists an initial node containing all integers $\{c_1,c_2,\dots, c_M\}$ in no specific order and none of the $M$ jobs is yet assigned. An example of BB as an SD technique to solve an assignment problem is shown using a state space tree in Fig. \ref{fig:branch-and-bound-tree-diagram} with $M = 4$. For simplicity, the initial node for $M = 4$ is $1234$ and nodes $1/234$, $2/134$ up to $4/123$ are obtained at $\epsilon = 1$. The node $t$ with the minimum cost $\hat{\delta}^{(t)}_{\epsilon}$ at $\epsilon = 1$ with the minimum value from the $M$ branches is scheduled if
\begin{equation} 
    \hat{\delta}_{\epsilon}^{(t)} = \min_{1 \leq t \leq M} (\hat{y}_{\epsilon t} + \sum_{i,j} \hat{y}_{ij}), \:\:\: \;\;\;\; \epsilon = 1, 2 \leq i \leq M, 0 < j \leq M: j \neq t. 
    \label{branch-and-bound-level-0}
\end{equation}
The other nodes are pruned while the surviving node $\hat{t}_{\epsilon}$ at $\epsilon = 1$ satisfying (\ref{branch-and-bound-level-0}) is divided into $M-\epsilon$ branches at the next level. At nodes $\epsilon > 1$, the surviving node at each level
\begin{equation} 
    \hat{\delta}_{\epsilon}^{(t)} = \min_{1 \leq t \leq M} (\hat{y}_{(\epsilon-1)\hat{t}_{(\epsilon-1)}} + \hat{y}_{\epsilon t} + \sum_{i,j} \hat{y}_{ij}), \:\:\: \;\;\;\; \epsilon > 1, \epsilon+1 \leq i \leq M, 1 \leq j \leq M : j \neq t, j \notin \boldsymbol{\hat{t}}, 
    \label{branch-and-bound-level-1}
\end{equation}
where $\boldsymbol{\hat{t}}$ contains an ordered set of scheduled nodes.
This process of pruning and branching is done until the last level at which only one assignment is possible.

\subsection{Schemes} \label{sec:schemes}
\begin{figure}
\centering\includegraphics[width=30pc]{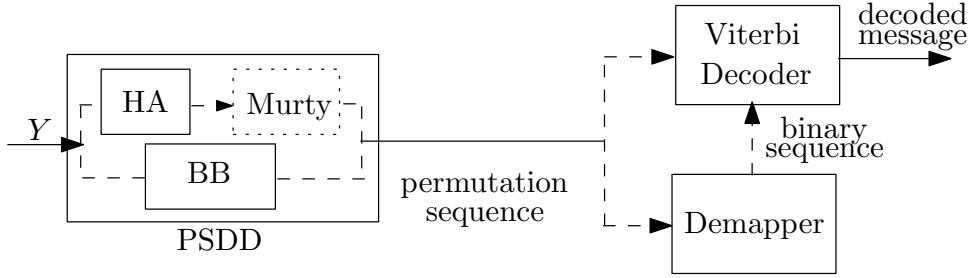}
\caption{Decoder system model of permutation code soft-decision and permutation trellis decoder.}
\label{fig:decoder-sd}
\end{figure}

As shown in Fig. \ref{fig:decoder-sd}, the PSDD uses BB or HA and Murty algorithms to decode $\boldsymbol{Y}$, hence the dashed lines. The dotted lines around the Murty block indicates the Murty algorithm is only used when $\boldsymbol{c} \notin \boldsymbol{C}$ at HA. Otherwise, the Murty step is skipped.

\subsubsection{Scheme 1}
The channel output $\boldsymbol{Y}$ is decoded using HA ($g = 1$) and Murty ($2 \leq g \leq M$). The codeword matrix which maximizes (\ref{eq:optimization-problem}) is then decoded using the Viterbi decoder. If HA produces a sequence $\boldsymbol{c} \notin \boldsymbol{C}$, Murty iterates until $\boldsymbol{c} \in \boldsymbol{C}$ or the specified maximum number of iterations is reached. 

\subsubsection{Scheme 2}
Here, the Viterbi decoder is used to decode the binary output from the permutation to binary code demapper. The PSDD produces a permutation code $\boldsymbol{c}$ that is demapped to its binary equivalent.
If the final iteration of the PSDD produces a codeword $\boldsymbol{c} \notin \boldsymbol{C}$, minimum distance decoding between $\boldsymbol{c}$ and all codewords in $\boldsymbol{C}$ is performed before demapping. This binary output is then decoded using the Viterbi decoder. 

\subsubsection{Schemes 3 \& 4}
In Scheme 3, the channel output $\boldsymbol{Y}$ or demodulated $\bar{\boldsymbol{Y}}$ is decoded with BB to solve for $Z_1$. The permutation sequence produced is then forwarded to the Viterbi decoder. In Scheme 4, the permutation sequence produced by BB is demapped to its binary equivalent and forwarded to the Viterbi decoder, similar to Scheme 2. BB uses a single iteration, hence only $Z_1$ is solved in Schemes 3 and 4.

In the case where the threshold detector is used at the demodulator, the SD algorithms can also be applied to demodulated bits. As an example, consider a transmitted codeword 3214. The demodulated codeword matrix $\bar{\boldsymbol{Y}}_{\text{I}}$ affected by impulse noise at time $t_4$ can be represented as
\begin{equation}
  \label{eq:received-y-job-worker}
  \bar{\boldsymbol{Y}}_{\text{I}} =
  \scalemath{1.0}{
  \begin{blockarray}{*{4}{c} l}
    \begin{block}{*{4}{>{$\footnotesize$}c<{}} l}
      t_1 & t_2 & t_3 & t_4 \\
    \end{block}
    \begin{block}{[*{4}{c}]>{$\footnotesize$}l<{}}
      0 & 0 & \fbox{1} & 1 \\
      0 & \fbox{1} & 0 & 1 \\
      \fbox{1} & 0 & 0 & 1 \\
      0 & 0 & 0 & \fbox{1} \\
    \end{block}
  \end{blockarray} } \text{ or } 
  \bar{\boldsymbol{Y}}_{\text{N}} =
  \scalemath{1.0}{
  \begin{blockarray}{*{4}{c} l}
    \begin{block}{*{4}{>{$\footnotesize$}c<{}} l}
      t_1 & t_2 & t_3 & t_4 & \\
    \end{block}
    \begin{block}{[*{4}{c}]>{$\footnotesize$}l<{}}
      1 & 1 & \fbox{1} & 1 & f_1 \\
      0 & \fbox{1} & 0 & 0 & f_2 \\
      \fbox{1} & 0 & 0 & 0 & f_3 \\
      0 & 0 & 0 & \fbox{1} & f_4 \\
    \end{block}
  \end{blockarray} },
\end{equation}
when affected by NBI at $f_1$.
$\bar{\boldsymbol{Y}}_{\text{I}}$ and $\bar{\boldsymbol{Y}}_{\text{N}}$ can be easily solved using the PSDD to produce the codeword 3214 as shown by the boxed values. The codeword can then be decoded using the Viterbi algorithm.

\section{Simulation Results and Complexity Analysis}
We compare the simulated BER of the PTC decoder with the BERs obtained from the proposed decoding schemes discussed in Section \ref{sec:schemes}. The BER is calculated as the ratio between the difference between the transmitted message bits and decoded bits, and the total number of transmitted bits. This is evaluated over an AWGN and PLC channel. Different PTC mappings are simulated, with three different codes. The first code has a rate $R_{\text{C}} = 1/2$ mapped onto $M = 3$, the second has a rate $R_{\text{C}} = 2/3$ mapped onto $M = 4$ codebook while the third is a distance preserving mapping from $R_{\text{C}} = 1/4$ onto an $M = 4$ codebook. The first is generated from $K = 3$, 
with an octal generator polynomial (7 5) while the second is generated from $K = 4$, 
with an octal generator polynomial (1 3 0; 3 2 3). The mappings of both codes have been shown in Table \ref{table:binary-to-permutation-mapping}. The third code is generated from $R_{\text{C}} = 1/4$, $K = 6$, 
with an octal generator polynomial (53 67 71 75). 
The third code is mapped onto a distance preserving permutation codebook of $M = 4$. 
The SNR per bit $E_b/N_0$ of the system for bit energy $E_b$ is defined as
\begin{equation} 
  \begin{aligned}
  \label{eq:snr-per-bit}
    \dfrac{E_b}{N_0} = \dfrac{E_s}{N_0 \times R_{\text{P}} \times \text{log}_{2}{M}}.
  \end{aligned}
\end{equation}
For the PLC channel, we adopt the heavily disturbed channel scenario in \cite{ma-so-gunawan} where the inter arrival time is 0.0196 s while $T_{\text{noise}}$ is 0.0641 ms. 

\begin{figure}
    \centering
    \includegraphics[width=20pc]{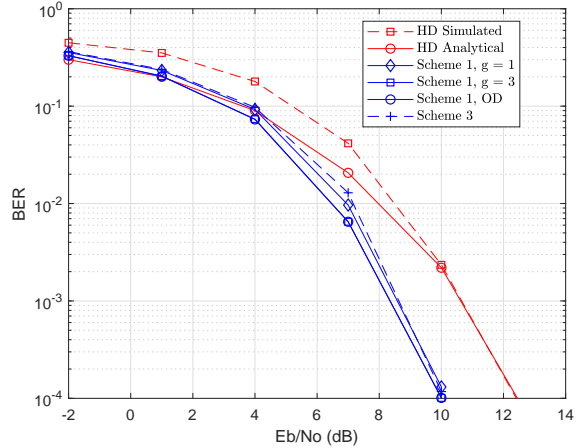}
    \caption{BER comparisons of HD with SD schemes for code having $R_{\text{C}}$ = 1/2 and $M$ = 3.}
    \label{fig:ptc-awgn-analytical-schems}
\end{figure}

\begin{figure}[!ht] 
  \begin{subfigure}[b]{0.5\linewidth}
    \centering
    \includegraphics[width=20pc]{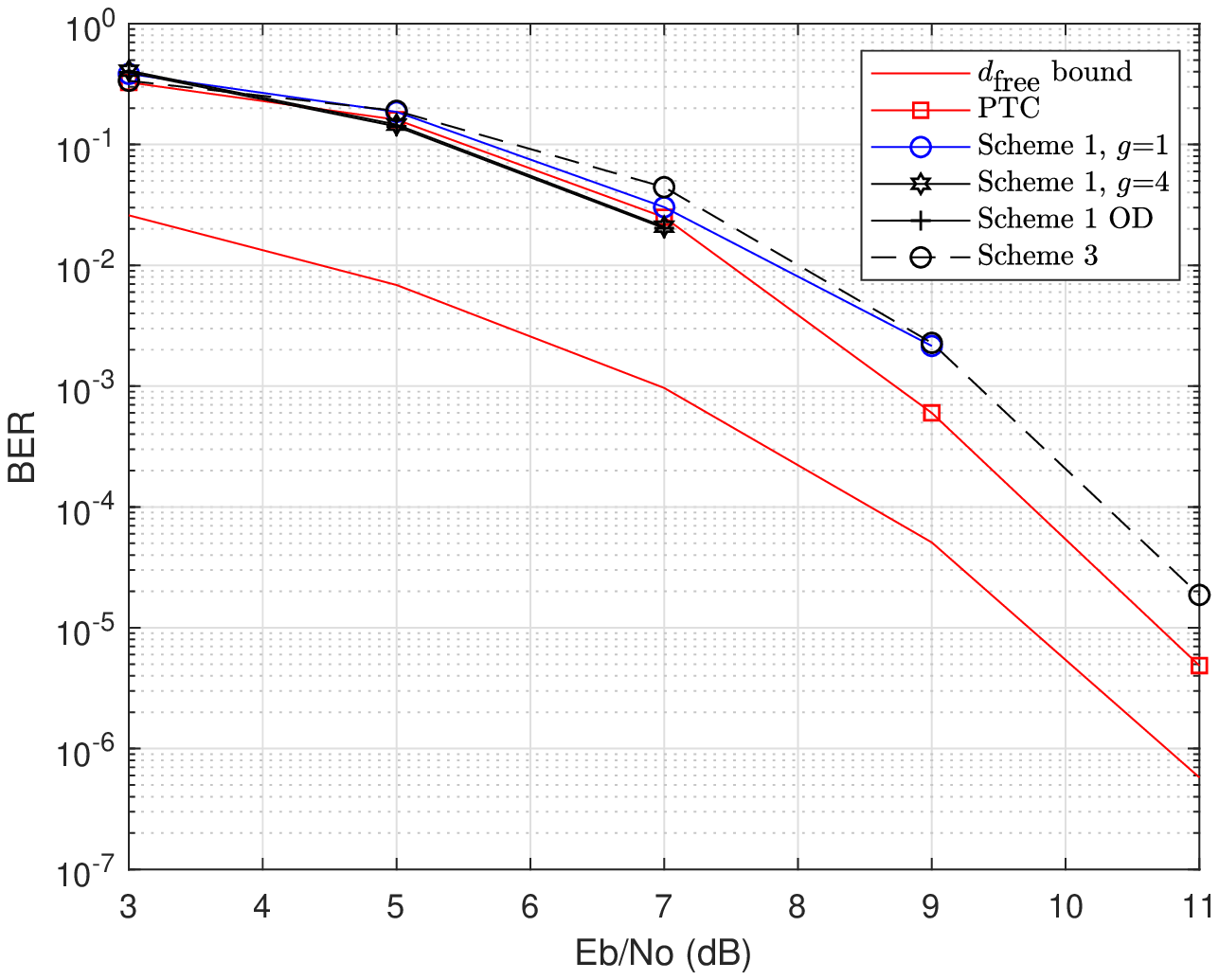}
    \caption{Schemes 1 \& 3 with Viterbi decoding of PTC codewords.} 
    \label{fig:ptc-awgn-123}
    \vspace{2ex}
  \end{subfigure}
  \begin{subfigure}[b]{0.5\linewidth}
    \centering
    \includegraphics[width=20pc]{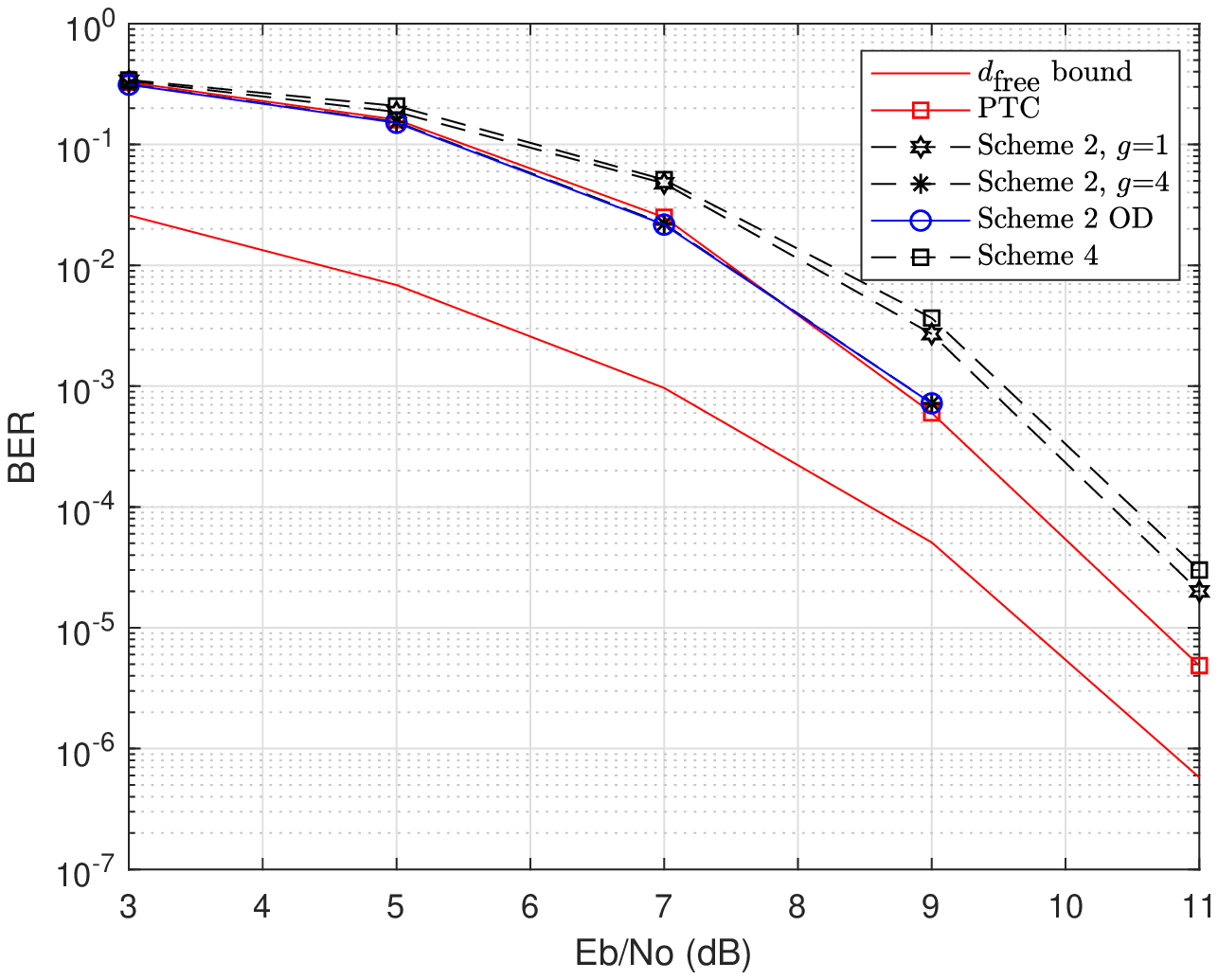}
    \caption{Schemes 2 \& 4 with Viterbi decoding of demapped codewords.} 
    \label{fig:ptc-awgn-demapped-123}
    \vspace{2ex}
  \end{subfigure}
  \caption{BER comparing the four PSDD decoding schemes with the PTC decoder for $R_{\text{C}}$ = 1/2 and $M$ = 3 in AWGN channel.}
\label{fig:half-rate-awgn}
\end{figure}

Figs. \ref{fig:ptc-awgn-analytical-schems} and \ref{fig:half-rate-awgn} show the performance of the SD schemes over the AWGN channel. The SD schemes are compared with the expressions in (\ref{eq:definition-of-P2d}) - (\ref{eq:definition-of-P-r-t}). The simulated HD plot matches the analytical plot at higher SNRs, while the SD decoders improve the coding gain by about 2 dB. The BB method in Scheme 3 shows a slightly poorer performance when compared with Scheme 1 at lower SNRs. 
Results in Figs. \ref{fig:ptc-awgn-123} and \ref{fig:ptc-awgn-demapped-123} show a tight BER performance between the modified PTC decoder and the SD schemes. The BER of the Viterbi decoder when the all-zero codeword is compared with the $d_{\text{free}}$ path, as described in (\ref{eq:pe-ptc}) is also shown in \ref{fig:ptc-awgn-123} and \ref{fig:ptc-awgn-demapped-123} as the $d_{\text{free}}$ bound. This bound can be considered as a lower bound. When $M$ is increased in Figs. \ref{fig:ptc-awgn-234} and \ref{fig:ptc-awgn-demapped-234}, the BER coding gains obtainable from Schemes 1 and 2 become more evident with up to 3 dB gain at $g = 4$. Figs. \ref{fig:ptc-plc-144} and \ref{fig:ptc-plc-demapped-144} show significant coding gains in the PLC channel with Schemes 3 and 4 performing close to the OD of Schemes 1 and 2.

\subsection{BER Performance and $\frac{2^n}{M!}$ ratio}
\begin{figure}[!ht] 
  \begin{subfigure}[b]{0.5\linewidth}
    \centering
    \includegraphics[width=20pc]{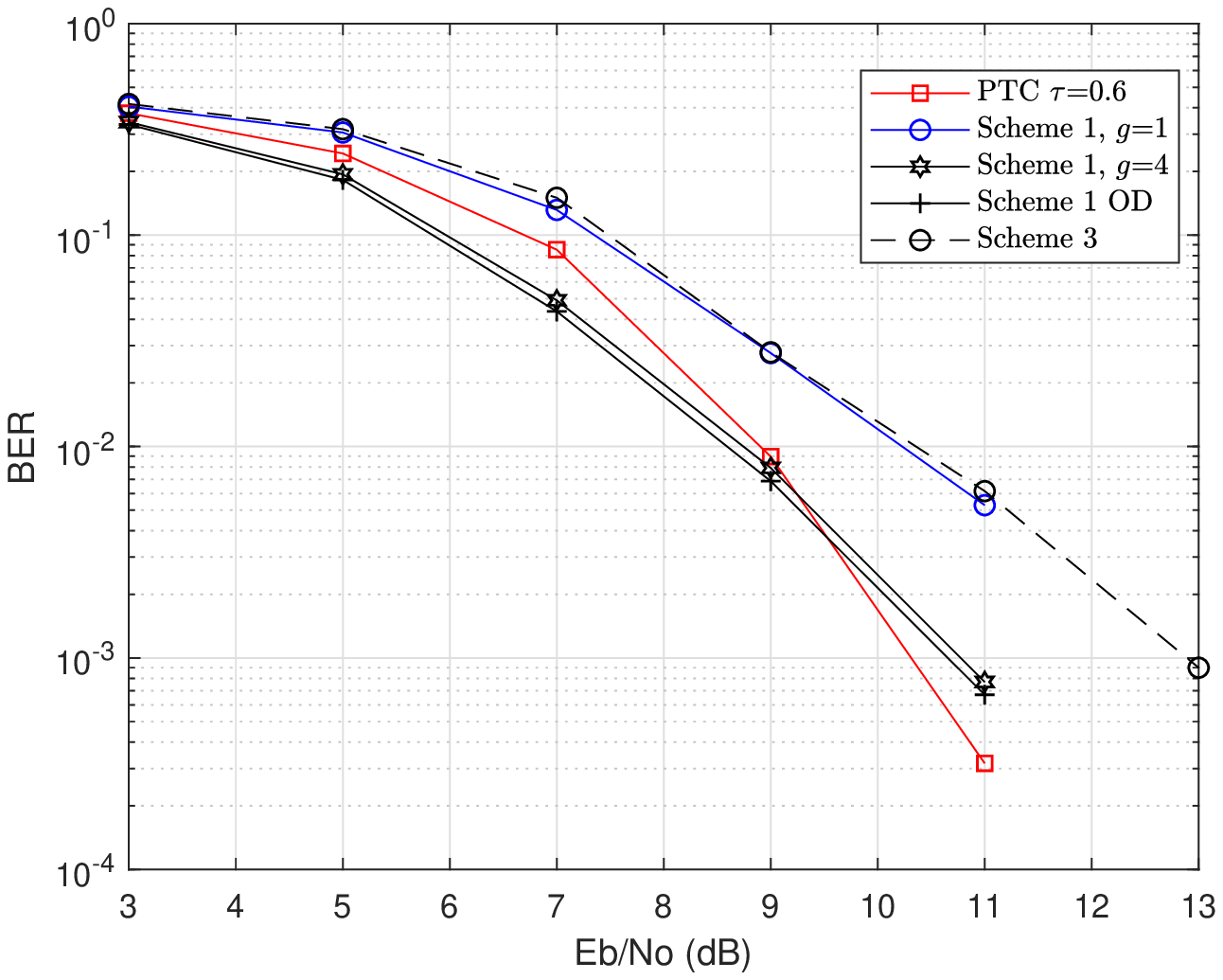}
    \caption{Schemes 1 \& 3 with Viterbi decoding of PTC codewords.} 
    \label{fig:ptc-awgn-234}
    \vspace{2ex}
  \end{subfigure}
  \begin{subfigure}[b]{0.5\linewidth}
    \centering
    \includegraphics[width=20pc]{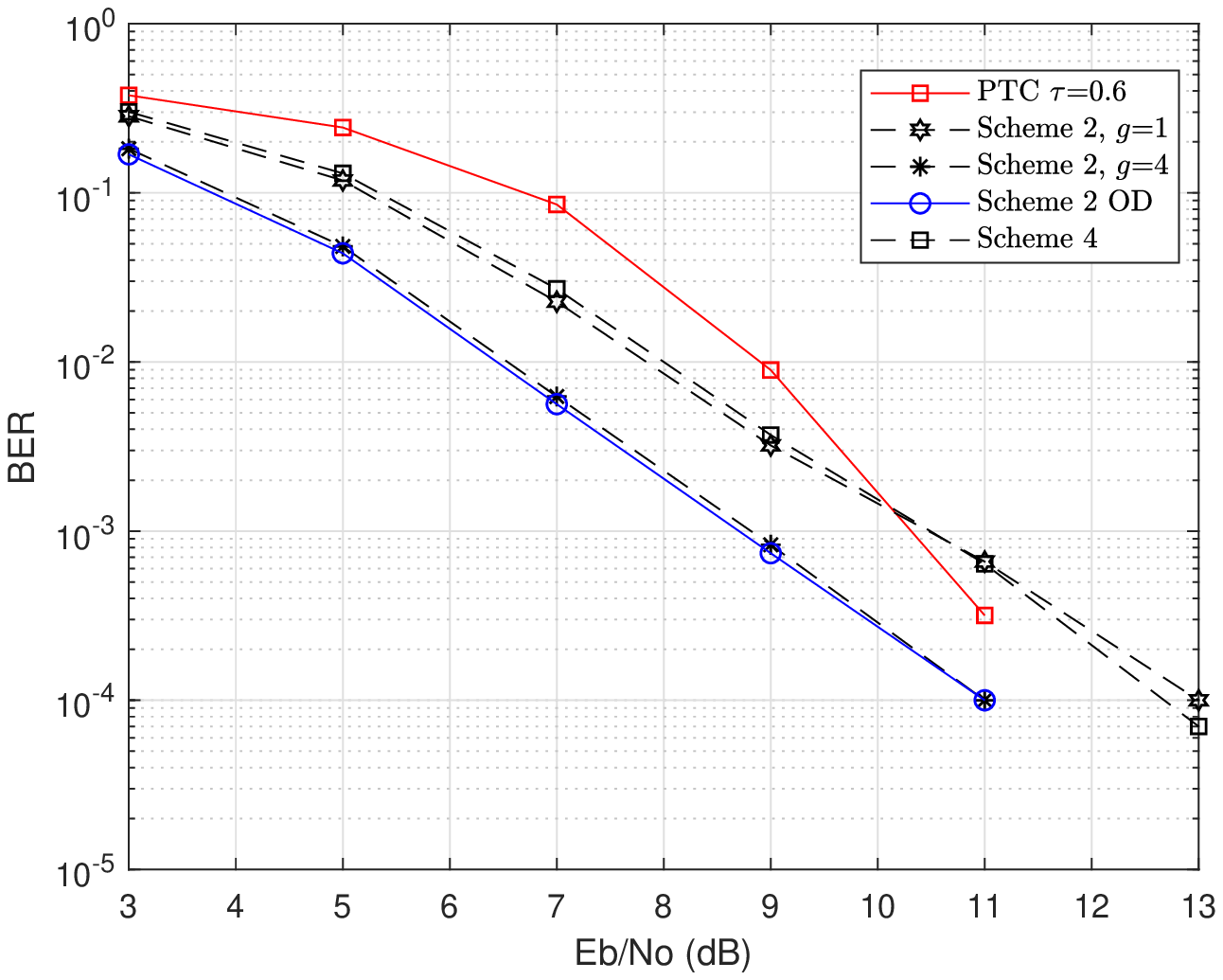}
    \caption{Schemes 2 \& 4 with Viterbi decoding of demapped codewords.} 
    \label{fig:ptc-awgn-demapped-234}
    \vspace{2ex}
  \end{subfigure}
  \caption{BER of PTC decoding for $R_{\text{C}}$ = 2/3, $M$ = 4 and $A$ = 0.1 in PLC channel.}
\label{fig:two-third-rate-awgn-iterations}
\end{figure}

\begin{figure}[!ht] 
  \begin{subfigure}[b]{0.5\linewidth}
    \centering
    \includegraphics[width=20pc]{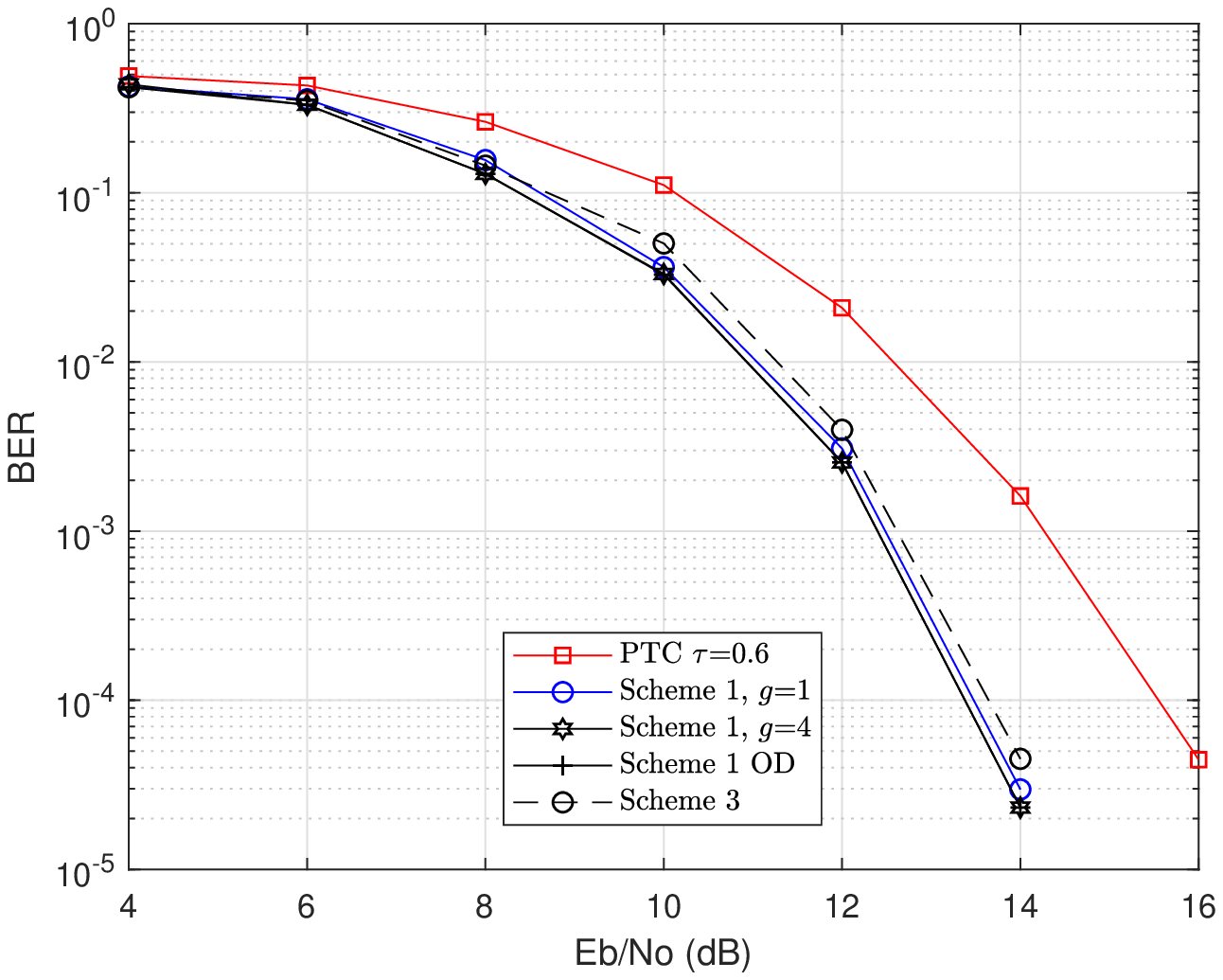}
    \caption{Schemes 1 \& 3 with Viterbi decoding of PTC codewords.} 
    \label{fig:ptc-plc-144}
    \vspace{2ex}
  \end{subfigure}
  \begin{subfigure}[b]{0.5\linewidth}
    \centering
    \includegraphics[width=20pc]{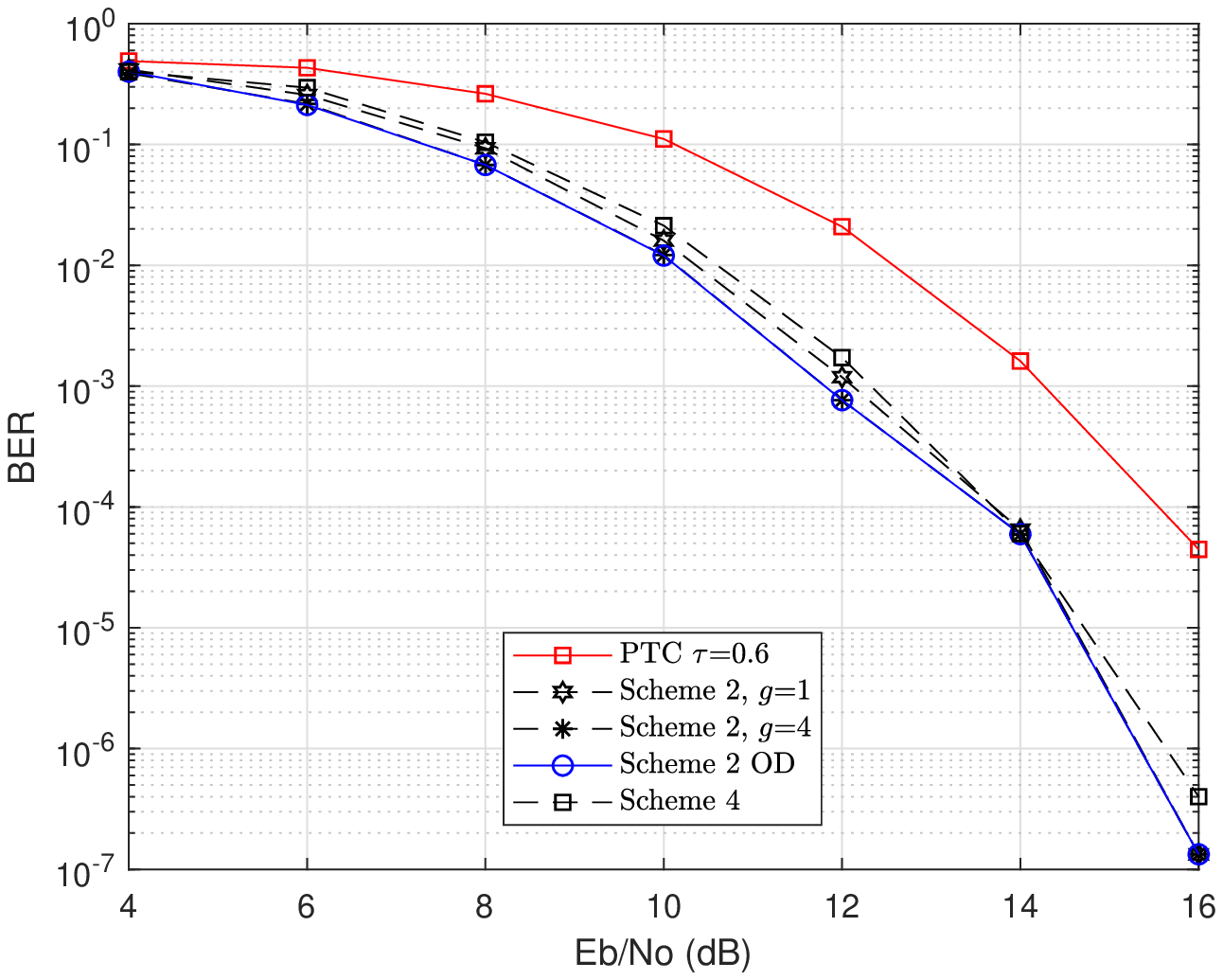}
    \caption{Schemes 2 \& 4 with Viterbi decoding of demapped codewords.} 
    \label{fig:ptc-plc-demapped-144}
    \vspace{2ex}
  \end{subfigure}
  \caption{BER of PTC decoding for $R_{\text{C}}$ = 1/4, $M$ = 4 and $A$ = 0.1 in PLC channel.}
\label{fig:one-quarter-rate-plc}
\end{figure}

The SD decoders presented provide the advantage of always outputting a permutation of integers by solving an assignment problem. The SD decoder's probability of producing a codeword $\boldsymbol{c} \in \boldsymbol{C}$ can be described by the ratio $\frac{2^n}{M!}$. Hence, the performance of the schemes approaches the OD performance as $2^n$ increases. For example, in the plots in Fig. \ref{fig:one-quarter-rate-plc}, the $2^{n}$ unique binary sequences are mapped onto a codebook with $M = 4$. Since $\frac{2^n}{M!} = 0.667$, each scheme of the SD decoders is more likely to choose a codeword $\boldsymbol{c} \in \boldsymbol{C}$. In Figs. \ref{fig:half-rate-awgn} and \ref{fig:two-third-rate-awgn-iterations}, more iterations such as $g = 4$ are required to produce $\boldsymbol{c} \in \boldsymbol{C}$ compared to $g = 1$. 
Schemes 2 and 4 further improve the performance of the SD because the demapping process chooses one of the $2^n$ possible binary sequences using Hamming distance decoding. In Schemes 1 and 3, the codeword matrix produced by the SD decoder likely belongs to $M!$ possible permutations, rather than the encoder's $2^n$ codewords due to errors. Since $2^n < M!$, the probability of error of the schemes using demapping is lower, especially at lower SNRs. 

\subsection{Complexity Analysis}
In order to analyze the different complexities of the decoder, operations required to process the digital signal are considered. The complexity of the circuitry required for the processing is not considered in this work. For each received, coded signal at the output of the channel, the TD demodulates each received sample of the $M \times M$ code matrix. This requires a complexity of $O(M^2)$ to decide if each sample in the $M \times M$ matrix is above or below the threshold. The demodulated $M \times M$ matrix is further compared with the codebook, to find the codeword having the minimum Hamming distance. This step will compare all the $2^n$ codewords with the demodulated matrix, having $M^2$ elements. This requires a complexity of $O(2^n \cdot M^2)$. For the PSDD, the complexity of the iterative decoder in Scheme 1 and 2 is $O(M^3)$ \cite{liu-shell} at $g = 1$ and $O(M^4)$ \cite{cox-miller-danchick-newnam} for $2 \leq g\leq M$. When $g > 1$, the complexities sum up, hence a worst case complexity of $O(M^4)$. At each level of the tree-based BB method in Fig \ref{fig:branch-and-bound-tree-diagram} and (\ref{branch-and-bound-level-0}), $M$ different computations are required on all the $M$ nodes in order to find the minimum cost at each level. Since the number of nodes to process reduce at each level, the worst case complexity is considered and can be approximated as $O(M^3\text{log}M)$ \cite{lageweg-lenstra-rinnooy-kan}. Schemes 2 and 4 require a lookup process to demap to the binary equivalent at a complexity of $O(M \text{log}M)$. 

\section{Conclusion}
The design of four decoding schemes, which combine three optimization algorithms to improve the BER performance of the PTC decoder is presented. The complexities of the different components of the decoding schemes are also analyzed. Demapping the permutation sequence to its binary equivalent before decoding with the Viterbi decoder produces the best BER coding gain when compared to the PTC decoder. The BER coding gains obtained from the schemes increase as the ratio of the permutation codebook's cardinality to all possible permutations increases. While the improved BER coding gain is at the expense of additional complexity, a reasonable trade-off can be considered when the permutation codebook contains large codewords. While multi-tone $M$-FSK has been established to improve the data rate of the $M$-FSK scheme, future work will require techniques for increasing the data rate of permutation-coded $M$-FSK.

\section*{Acknowledgement}
The authors would like to thank and acknowledge the financial support provided by South Africa's National Research Foundation (112248 \& 114626) and the Sibanye-Stillwater Digital Mining Laboratory (DigiMine), Wits Mining Institute (WMI), University of the Witwatersrand, Johannesburg, South Africa.

\bibliography{main}
\bibliographystyle{elsarticle-num}

\end{document}